\documentstyle[12pt]{article}

\begin{document}
\begin{flushright}
November 1994
\end{flushright}
\begin{center}
{\Large A Nonperturbative, Schwinger-Dyson-Equation Analysis of Quark
Masses and Mixings in a Model with QCD and Higgs Interactions}
\end{center}
\vspace*{.5cm}
\begin{center}
{\Large L}ESLEY {\Large L}. {\Large S}MITH$^{a}$, {\Large P}ANKAJ
{\Large J}AIN$^{b}$,
{\Large D}OUGLAS {\Large W}. {\Large M}C{\Large K}AY$^{a,c}$
\vspace{.5cm}
\newline  $^{a}$ Department of Physics and Astronomy, University of Kansas,
Lawrence, KS 66045
\newline  $^{b}$ Department of Physics and Astronomy, University of Oklahoma,
Norman, OK, 73019
\newline  $^{c}$ Kansas Institute for Theoretical and Computational Science,
University of Kansas, Lawrence, KS 66045
\end{center}
\vspace*{.5cm}
\begin{center}
{\bf Abstract}
\end{center}
The Schwinger-Dyson equation for the quark self-energy
is solved in the quenched ladder approximation for several cases of one-
and two-quark-generations.  The exchanges of standard model gluons and Higgs
bosons are taken into account.  It is found that Higgs boson exchange
dominates the quark self-energy in the ultraviolet region
for sufficiently large input quark masses ($>75$ GeV),
causing the running quark propagator mass to increase with energy-scale.  The
running of the quark mixing angles is also considered.  No running of the
quark mixing angles is found for input
quark masses up to and including 500 GeV.
\pagebreak
\section*{I. Introduction}
\indent
\par
Unlike the renormalization group equation,
the Schwinger-Dyson equation
 (SDE) analysis presented here
enables us in principle to calculate the running quark mass functions
for very light quarks at low energies
where quantum chromodynamics (QCD) is in the non-perturbative region.
This is of interest because
experimentally accessible quantities are presently found at low
energy-scales.  In the presence of heavy quarks, such as bottom and top,
it allows the calculation of running mass matrix
at momentum scale of the order or less than the quark masses, at which
scale the RG equation results may not be reliable.
Moreover,
 this nonperturbative SDE approach is also interesting in that it can
describe the running mass function of an ultra-heavy quark, with a large
 Yukawa coupling that cannot be treated reliably by perturbation
theory.  This would be relevant for a fourth
generation quark, for example.
\par
The calculation of the variation of mixing angles with momentum
scale is of great experimental interest since they are usually measurable
only in a very limited kinematic region.
  To illustrate this point, consider
the top-strange CKM mixing parameter, $V_{ts}$.  Direct top-antitop quark
bound state production at the next linear collider would occur at
approximately $> \, 300 \,{\rm GeV}^{2}$,
and $V_{ts}$ would be evaluated there.  Presently, some estimations
of $V_{ts}$ are being done indirectly using penguin diagrams
in the calculation of $b \rightarrow s$ transitions.  In this method
a virtual top quark mixes with a virtual strange quark at a few GeV$^{2}$
\cite{CLEO94}.
Clearly if
$V_{ts}(300\; {\rm GeV}^{2})\, \neq \, V_{ts}(\simeq\; 3\; {\rm GeV}^{2})$,
complications would arise.
This possibility should be investigated both by
RG techniques and by the SDE. In particular SDE studies
can reveal
the existence of any nonperturbative effects which might be absent in RG
studies.
%
\par
There have been several studies of the quark propagator Schwinger-Dyson
Equation.  These have fallen into the areas of
chiral-symmetry breaking \cite{NJL,Peskin,Fomin,AJ,chiSB}, confinement
\cite{Pagels76}, and related phenomenology
including pseudoscalar decay constants and form factors
\cite{Pagels79,Jackiw,MM89,Barducci}.
The complexity of the SDEs makes it neccesary to use various approximations
or assumptions.  The numerous QCD quark SDE studies
make assumptions about the quark-gluon vertex and they
model the effective gluon propagator.  Many of these studies
utilize the Landau gauge, ladder approximation and the
angle approximation to express the integral SDE in terms of
a differential equation \cite{RW94}.
Recent reviews of SDE literature
which include discussion of and references for the above points
include \cite{RW94,Hadicke,Roberts93}.
%
\par
Though the propagator is a gauge dependent object, and this is true in the
approximations that we use,
 our results do agree with
the gauge-invariant one-loop renormalization group analysis results in the
ultraviolet region.
We are careful not to neglect derivatives of the strong coupling,
$\alpha_{s}(q^{2})$, to be consistent with the one-loop
RGE result \cite{Higash}.
\par
Roberts and McKellar \cite{RM90} have studied the validity
of the angle approximation used in the context of the model of
Atkinson and Johnson \cite{AJ}.  They use the
Landau gauge and the running strong coupling
$\alpha_{s}(q^{2})=d \pi /ln(x_{0}+q^{2}/\Lambda^{2}_{QCD})$, and re-express
the integral equation as a differential equation.  Roberts and McKellar find
that the angle approximation is qualitatively reasonable,
introducing errors of order 12\%.  For other models, however,
the angle approximation is not so reliable, particularly when the strong
coupling becomes singular at the origin.
To minimize problems with the angle approximation,
we utilize some aspects of the model of
Atkinson and Johnson \cite{AJ}.
\par
Rather than investigate chiral symmetry breaking,
this study
assumes that chiral symmetry is broken and
solves the quark propagator SDE to find running quark mass functions
and running quark mixing angles.  In addition to the usual gluon
contributions, we include Higgs boson exchange effects
in the quark self-energy.  The
contributions to the quark self-energy due to Higgs bosons are being studied
here for the first time in the context of the quark propagator SDE.  We also
study for the first time in a Higgs-boson-plus-gluon model the
multigenerational
cases of quarks which enable us to analyze quark mixing angles.
%
%
\section*{II. Model}
\indent
\par
The general form of the quark propagator SDE that we use is
\begin{equation}
S^{-1}(q)=S^{-1}_{0}(q)-\int \frac{d^{4}k}{(2\pi)^{4}} g_s\gamma_{\mu}
S(k)g_{s}\Lambda_{\nu}G^{\mu\nu}-\int \frac{d^{4}k}{(2\pi)^{4}} g_{H}
S(k)g_{Y}\Lambda_{H}P_{H}, \label{SDE}
\end{equation}
where $g_s\gamma_{\mu}$ and $g_{s}\Lambda_{\nu}$ are quark-gluon vertices,
$S(k)$ is the quark propagator, $G^{\mu\nu}$ is the gluon propagator,
$g_{Y}\Lambda_{H}$ is the quark-Higgs boson vertex factor, and $P_{H}$
is the Higgs boson propagator, to be discussed below.
\par
In this study we will simplify the quark propagator by using the ladder
approximation, namely we will set all the vertices to be the
bare vertices.  We will, however, use the RG improved expressions for
both the QCD and Yukawa couplings.
 In the Landau gauge $A(q)=1$ in the approximate form used here
of the Schwinger-Dyson-Equation (SDE) in QCD \cite{Kondo}.  Thus the ladder
approximation quark propagator in the Landau gauge in Minkowski space that
we use is $iS(k)=({\not}{k}+M)(k^{2}-M^{2})^{-1}$, where $M$ is a
$2\times2$ quark mass function matrix, assumed to be symmetric for this
study.  The ladder approximation quark
propagator is accurate enough for our purposes because this is an
exploratory study of a quark self-energy SDE which includes QCD and Yukawa
contributions \cite{note2}.
\par
The gluon propagator we use is
\begin{equation}
g^{2}_{s}G^{\mu\nu}(k) \equiv i\left(-g^{\mu\nu}+
\frac{k^{\mu}k^{\nu}}{k^{2}} \right)
G(k)
\end{equation}
with
\begin{equation}
G(k)=\frac{g^{2}_{s}(k)}{k^{2}}
\end{equation}
and the leading-log (one-loop) coupling
\begin{equation}
g^{2}_{s}(k)=
\frac{4\pi^{2}d}{ln(x_{0}-\frac{k^{2}}{\Lambda^{2}_{QCD}})}
\label{gQCD}
\end{equation}
where $d=12/(33-2n_{f})$, $n_{f}$ is the number of flavors.
This form has the correct $-k^{2} \rightarrow \infty$ leading order
QCD behavior (for $-k^{2}>0$).
This is a simple extension of the form used successfully to fit $J/\psi$
and $\Upsilon$ spectroscopic data \cite{Richardson}.  This leading-log QCD
coupling is renormalization-scheme and gauge independent, as is its second
order, two-loop, extension \cite{twoloop}.  $x_{0}$ is a parameter
that functions as a smooth infrared cutoff.  Equation (\ref{gQCD}) does not
accurately model the gluon potential in the infrared
region, but it does enable
us to assess the relative magnitude and shape effects
of the Higgs boson and gluon exchanges in the quark self-energy \cite{note3}.
\par
The QCD coupling also contains the parameter $n_{f}$, the number of
quark flavors.
The number of quark flavors is determined by the energy scale,
$y=\frac{q^{2}}{\Lambda^{2}_{QCD}}$, where $q$ is the momentum transfer.
We use a step-function expression to
increment $n_{f}$ at $q=$1.27 GeV ($n_{f}=4$),
 4.25 GeV ($n_{f}=5$),
160 GeV ($n_{f}=6$), and 500 GeV ($n_{f}=8$),
 where the running quark mass values
$m_{c}(m_{c})=1.27$ GeV, and $m_{b}(m_{b})=4.25$ GeV are from Gasser and
Leutwyler \cite{GL}.
%
\par
We must also model the Yukawa couplings.  We consider a $2\times2$ symmetric
Yukawa coupling matrix, which can be diagonalized by a $2\times2$ unitary
matrix, to give diagonal elements $g_{\pm}$.  The one-loop Renormalization
Group Equation for the diagonalized Yukawa couplings, $g_{\pm}$, without
weak interaction effects, is
\begin{equation}
\frac{dg_{\pm}}{dt}=\frac{1}{4\pi^{2}}
\left( \frac{9}{8}g^{3}_{\pm}(t)-2g^{2}_{QCD}(t)g_{\pm}(t) \right)
\end{equation}
with $t=ln(x_{0}+q^{2}/\Lambda^{2}_{QCD})/2$.  This has solution
\begin{equation}
g^{2}_{\pm}=\frac{1}{C_{\pm}t^{2d}-\frac{9}{16\pi^{2}}\frac{t}{1-2d}}
\label{gpm}
\end{equation}
where the $C_{\pm}$ must be determined by boundary conditions.
We apply the boundary
conditions
$m^{2}_{\pm}(1\; {\rm GeV})=g^{2}_{\pm}(1\;{\rm GeV}){\rm v}^{2}/2$
for small ($<150$ GeV) running quark masses, and
$m^{2}_{\pm}(m_{\pm})=g^{2}_{\pm}(m_{\pm}){\rm v}^{2}/2$ for large
running quark
masses, where v $\simeq$ 246 GeV, and $m_{+}=m_{t}$ and $m_{-}=m_{c}$,
if we consider the $2\times2$ mass matrix to represent the second and
third generations,
for example.  We chose $\Lambda_{QCD}=0.18$ GeV to set the boundary values of
$g_{\pm}$.
\par
We convert Eq. (\ref{SDE}) into 4-d spherical polar
coordinates, and integrate over $\phi$ and $\chi$ analytically.  To
 integrate over $\theta$, we employ the so-called angle
approximation \cite{Higash} in the QCD sector and its analog in the Yukawa
sector.
The angle approximation is equivalent to keeping the first term in an
expansion of the angular, $\theta$, integral \cite{MM90}.  It predicts the
same leading behavior for the ultraviolet asymptotic forms of M(y) as given by
operator-product-expansion analysis in the pure QCD case
\cite{Politzer}.
Roberts and McKellar \cite{RM90} showed that the angle approximation may be
useful
for a qualitative study of the quark SDE when the infrared behavior
is expected to
be smooth, as in our gluon model.
We also make the approximation
$M^{2}_{H} \ll \Lambda^{2}_{QCD}(x+y-2\sqrt{xy}cos\theta)$.
This approximation is appropriate for the heavy quark cases that we study.
For the light quark cases, this approximation exaggerates Higgs effects,
thereby
providing an upper bound to any new nonperturbative effects.
For all input quark masses this Yukawa sector approximation holds in the
asymptotic region where $y \gg 10^{5}$.
\par
The resulting 1-d SDE is thus
\begin{displaymath}
M=M_{0}+\Lambda^{2}_{QCD}\int^{y}_{0}xdx \left[-d\,M\cdot(-x\Lambda^{2}_{QCD}
-M^{2})^{-1}\cdot \frac{1}{yln(y+x_{0})}
\right.
\end{displaymath}
\begin{displaymath}
\left.
+\frac{1}{32\pi^{2}}g_{Y}(y)\,M\cdot(-x\Lambda^{2}_{QCD}-M^{2})^{-1}
\frac{1}{y} g_{Y}(y) \right]
\end{displaymath}
\begin{displaymath}
+\Lambda^{2}_{QCD} \int^{\infty}_{y} x dx \left[ -d\, M\cdot
(-x\Lambda^{2}_{QCD}-M^{2})^{-1} \cdot \frac{1}{xln(x+x_{0})}
\right.
\end{displaymath}
\begin{equation}
\left.
+\frac{1}{32\pi^{2}}g_{Y}(x)\,M\cdot(-x\Lambda^{2}_{QCD}-M^{2})^{-1}
\frac{1}{x} g_{Y}(x)\right]. \label{SDE1d}
\end{equation}
\par
We solve Eq. (\ref{SDE1d}) by converting it into a second order differential
equation (see Appendix) and applying a fourth-order Runge-Kutta subroutine.  To
ensure that the differential equation is equivalent to our integral
equation, we must enforce some initial conditions.
The QCD quark propagator integral SDE relates the values for the derivatives
of the quark masses at the origin to the quark masses themselves.  For the
pure QCD 1-particle case,
$\frac{dM}{dy}|_{y=0}=-\frac{\alpha_{QCD}(0)}{2M(0)\pi}$, where
$\alpha_{QCD}(0)=\pi d/ln(x_{0})$ \cite{MM90}.  The initial conditions for
our QCD plus Higgs boson study must be consistent with this.  For the
two-quark-generation case, we assume the propagator
quark mass function takes the
form $M_{ij}(y)=M_{0\,ij}+M_{1\,ij}\, y$ near the origin, where $M_{0ij}$
are the initial input quark masses, $M_{q}(0)$, and the forms
of the first derivatives of the mass matrix elements, $M_{1\,ij}$,
are given in the Appendix.
The pure QCD sector tells us
further that since we have included a nonzero
bare quark mass (see Eq. \ref{SDE1d}),
we must agree with the irregular QCD mass solution in the
asymptotic region, i.e. for $y \gg 1$, $M(y)\propto (ln(y+x_{0}))^{-d}$
as long as QCD is the dominant effect \cite{AJ}.  We will use this as an
asymptotic test of our numerical results.
\par
\begin{center}
\begin{tabular}{|c||c|c|} \hline
quark & $M_{q}$(0 GeV) & $m_{q}$(1 GeV) \\ \hline\hline
up & 560 MeV & 5.6 MeV \\ \hline
down & 560 MeV & 9.9 MeV \\ \hline
strange & 720 MeV & 199 MeV \\ \hline
charm & 1.66 GeV & 1.32 GeV \\ \hline
bottom & 4.89 GeV & 4.52 GeV \\ \hline\hline
\end{tabular}
\end{center}
\begin{center}
{\bf Table 1a:  Input quark masses}
\end{center}
\pagebreak
\par Numerical input values are given in Tables 1.
\begin{center}
\begin{tabular}{|c||c|c|} \hline
quark & $M_{q}$(0 GeV) & $m_{q}(m_{q})$ \\ \hline\hline
top & 179 GeV & 174 GeV \\ \hline
bottom$'$ & 506 GeV & 500 GeV \\ \hline
top$'$ & 506 GeV & 500 GeV \\ \hline\hline
\end{tabular}
\end{center}
\begin{center}
{\bf Table 1b:  Input quark masses}
\end{center}
We use two sets of input quark masses. For up, down, charm, strange, and
bottom quarks at q=0 we use Jain and Munczek's \cite{Munczek} values
as initial conditions on the $M_{q}(0)$ for the integration subroutine. The
1 GeV values are used to fix the values of $C_{\pm}$ of the Yukawa sector
(except for the top quark and fourth generation quarks).
The light quark (u,d,s) 1 GeV input masses are obtained by choosing values at
1 GeV which are consistent with the 1994 Particle Data Book and Y. Koide's
paper ``Table of Running Quark Masses" \cite{Koide94}.
The heavy quark (c and b) 1 GeV mass values are obtained by running the heavy
quark mass values given in the 1994
Particle Data Book to 1 GeV \cite{PDB94}.
The Schwinger-Dyson evolution of $M(q)$ is small below $m_{q}(q)$ for heavy
quarks \cite{Munczek}, as seen from the slightly higher evolved values of
$M_{t,t',b'}(0)$ starting from $m_{q}(m_{q})$ values.
For the top quark and
the fourth generation quarks we normalize $C_{\pm}$ at $m_{\pm}(m_{\pm})$
as indicated after Eq. (\ref{gpm}).
\par
For the multiple-quark-generation case we must also input the
$q^{2}/\Lambda^{2}_{QCD}=0$ values of the mixing angles
for the up-sector and down-sector.  We somewhat arbitrarily choose the
up-sector mixing angle to be $\theta_{up}=0.5$ radians.  To ensure that the
cabibbo mixing angle agrees with the accepted SM value for the first two
generations, $\theta_{cabibbo}=0.22$ radians at small $q^{2}$
\cite{PDB94}, we choose
$\theta_{down}=\theta_{up}-\theta_{cabibbo}$.  For the third-fourth
two-generation case, our convention is to adopt the same
$\theta_{cabibbo}$ input value, since it is unknown.
For the second-third two-generation case, we utilize the leading
angle expansion term of the 1994 Particle Data Book's $V_{cb}$,
$sin \theta_{23} \equiv 0.04$ \cite{PDB94}.
We are primarily interested in the question
of the running of the mixing angles, not their specific values, in any case.
Next we turn to the results of this analysis.
\section*{III. Results}
\indent
\par
We solve the one-quark-generation version of Eq. (\ref{SDE1d}) for
seven cases of quark input masses given in Table 1 for the running
quark mass functions.  We summarize some
characteristics of these
results, including the initial and final values of the quark mass
functions in Table 2, where $A$ and $B$ are the initial and final
integration points, for example $5\times 10^{-5}$, and
$9.77\times10^{8}$, respectively, in units of $\Lambda^{2}_{QCD}$.
We also compare the QCD plus
Higgs boson exchange case
with the QCD-only case in the asymptotic region to get the ratio
$\frac{M_{QCD+H}}{M_{QCD}}$.
\begin{center}
\begin{tabular}{|c||c|c|c|} \hline
quark & $M_{q}(A)$ & $M_{q}(B)$ & $\frac{M_{QCD+H}}{M_{QCD}}$ \\ \hline\hline
up & 3.11$\Lambda_{QCD}$ & 1.00$\Lambda_{QCD}$ & 1 \\ \hline
down & 3.11$\Lambda_{QCD}$ & 1.00$\Lambda_{QCD}$ & 1 \\ \hline
strange & 3.99$\Lambda_{QCD}$ & 1.42$\Lambda_{QCD}$ & 1 \\ \hline
charm & 9.22$\Lambda_{QCD}$ & 4.21$\Lambda_{QCD}$ & 1 \\ \hline
bottom & 27.2$\Lambda_{QCD}$ & 15.3$\Lambda_{QCD}$ & $\simeq$1 \\ \hline
top & 978$\Lambda_{QCD}$ & 825$\Lambda_{QCD}$ & 1.01 \\ \hline
bottom$'$=top$'$ & 2790$\Lambda_{QCD}$ & 3140$\Lambda_{QCD}$ &
$\neq$1, varies \\ \hline\hline
\end{tabular}
\end{center}
\begin{center}
{\bf Table 2: One-quark-generation results, for $A=5 \times 10^{-5}$,
$B=9.77\times10^{8}$}
\end{center}
\par
We solve the second order differential equation version of Eq. (\ref{SDE1d}),
with initial conditions given in an Appendix, to get the quark mass
function
weak eigenstates $M_{11}$, $M_{12}$, $M_{22}$.  We get the
 quark mass function eigenstates from
$M_{\pm}=\frac{1}{2} \left[ M_{11}+M_{22}\pm \sqrt{(M_{11}-M_{22})^{2}
+4M^{2}_{12}}\right]$, and the mixing angles from
$tan2\theta_{f}=2M_{12}[(M_{22}-M_{11})\pm\sqrt{(M_{22}-M_{11})^{2}
+4M^{2}_{12}}]^{-1}$, where $f$ stands for the flavor-sector, and where
$\theta_{cabibbo}=\theta_{up}-\theta_{down}$.  We look at three cases of
two-quark-generations: first and second generations, second and third
generations, and third and fourth generations.
\par
In Figure 1, we graph the bottom quark mass function for the
QCD plus Higgs boson case (solid line) and for the QCD-only case (dashed
line) from the second and third generations analysis.  The two lines are
coincident.  We also graph the top quark mass function for our
QCD plus Higgs boson model (solid line)
 from the second and third generations analysis in Figure 2, with the
corresponding QCD-only case (dashed line) shown for comparison.
  The QCD
plus Higgs boson case of the top quark mass function shows some deviation
from pure QCD.
The Higgs boson term has an opposite sign relative to the gluon term, so it
drives the mass function up, in contrast to the usual QCD-only decreasing
 mass function.
In Figure 3 we illustrate the dependence of the fourth generation
top$'$ quark mass function on the energy scale, from the
third and fourth generations case.  The Higgs boson has a
substantial effect on the fourth generation quark self-energy, i.e.
the QCD plus Higgs boson case (solid line) disagrees substantially with the
pure QCD case (dashed line).
In fact, the running top$'$ quark mass is increasing in the asymptotic
region.
In Figure 4
we graph the `running' Cabibbo mixing angle versus the energy scale for the
third-fourth generations case, which shows the largest Higgs effect
on the mass.  There does not appear to be any variation
in the mixing angle with respect to energy scale.
This is true in all cases we study.
\par
In Table 3 we summarize the two-quark-generation results.  The two-quark-
generation results are very similar to the one-quark-generation results,
as we see by comparing Tables 2 and 3.  Apparently the mixing effects
in our model in
the two-generation cases do not significantly affect the quark mass functions.
\begin{center}
\begin{tabular}{|c||c|c|c|} \hline
quark & $M_{q}(A)$ & $M_{q}(B)$ & $\frac{M_{QCD+H}}{M_{QCD}}$ \\ \hline\hline
up & 3.11$\Lambda_{QCD}$ & 1.00$\Lambda_{QCD}$ & 1 \\ \hline
down & 3.11$\Lambda_{QCD}$ & 1.00$\Lambda_{QCD}$ & 1  \\ \hline
strange & 3.99$\Lambda_{QCD}$ & 1.42$\Lambda_{QCD}$ & 1 \\ \hline
charm & 9.22$\Lambda_{QCD}$ & 4.21$\Lambda_{QCD}$ & 1 \\ \hline
bottom & 27.2$\Lambda_{QCD}$ & 15.3$\Lambda_{QCD}$ & 1 \\ \hline
top & 993$\Lambda_{QCD}$ & 840$\Lambda_{QCD}$ & 1.01 \\ \hline
bottom$'$=top$'$ & 2810$\Lambda_{QCD}$ & 3150$\Lambda_{QCD}$ &
$\neq$1, varies \\ \hline\hline
\end{tabular}
\end{center}
\begin{center}
{\bf Table 3:  Two-quark-generation mass results, for
$A=5 \times 10^{-5}$, $B=9.77 \times 10^{8}$.}
\end{center}
%
\section*{IV. Conclusions}
\indent
\par
All calculational results have been verified by an independent calculation
as indicated in the Appendix.
We have calculated the momentum-dependent, quark propagator mass for seven
flavors of one-quark-generation quarks via a nonperturbative SDE
treatment with Higgs boson exchange contributions.  We do not see the
effect in the ultraviolet region
on the quark self-energy
due to adding the Higgs boson interaction until the
input quark mass is sufficiently large, $M_{q}(0\;{\rm GeV})>75$ GeV.
Thus only the top quark and the fourth generation quark results differ from
pure QCD results.
\par
We have also calculated the quark mass function for eight flavors of
two-quark-generation quarks via a nonperturbative Schwinger-Dyson
matrix equation analysis with Higgs boson exchange contributions to
the quark self-energy.  The results are similar to the one-quark-generation
results.  Only the top quark and fourth generation quarks, top$'$, and
bottom$'$, differ significantly from the pure QCD results, with the
fourth generation quarks showing the most marked effect of the Higgs.
For the fourth generation quarks, the quark mass function actually
increases with energy-scale, due to Higgs boson exchange.
\par
We can analytically determine where the Yukawa term dominates.  In terms of
$y=q^{2}/\Lambda^{2}_{QCD}$ and
$d=12/(33-2n_{f})$ we get
\begin{displaymath}
y > \left| exp \left\{ \left[ \left(1-\frac{9d}{2d-1} \right)
\frac{(2)^{2d}}{32\pi^{2}dC_{\pm}} \right]^{\frac{1}{2d-1}} \right\} -x_{0}
\right|
\end{displaymath}
where $C_{\pm}<0$ for
$y>-x_{0}+exp\{ \frac{16\pi^{2}{\rm v}^{2}}{m^{2}_{\pm}} \frac{2d-1}{d} \}$.
Thus for the top quark, $C_{t}=-0.19$, and the Yukawa term dominates when
$q$ is beyond the Planck scale.  For the fourth
generation quark, $C_{4th}=-0.053$, and the Yukawa term dominates when
$q>1.5$ TeV.
\par
We have calculated the running mixing angle, which we call theta-cabibbo,
for three cases of two-quark-generations.
We detect no running of theta-cabibbo or any other mixing angle due to any
effect, for any quark mass input value up to and including 500 GeV.
This is expected for pure QCD and can be proven analytically \cite{Smith}.
This
provides a check of the numerical work.  By studying the mixing angle results,
we determine that
the running of the mixing angles due to nonperturbative effects
in our model is less than one part in $10^{9}$, far less than any
conceivable experimental sensitivity.  We have been unable to prove
analytically that the cabibbo angle is constant when the Yukawa interaction is
included in the SDE analysis, but we have established numerically
that there can be no observable effect even for the heavy hypothetical fourth
generation.  Based on our results, it appears to be safe to ignore running
of $V_{ts}$ and $V_{tb}$,
for example, in extrapolating from $m^{2}_{b}$ to $m^{2}_{t}$
in parametrizing data \cite{CLEO94,note1}.  Our nonperturbative results
agree in this respect with the RGE analysis of the model that we study,
since the RGE's can always be diagonalized completely and no running of the
mixing angles occurs.  As remarked earlier, the $y$-dependence of our mass
functions agrees with that of the 1-loop RGE's in the asymptotic region,
though not in the region at or below the scale of the input masses.
%
\par
In our investigation we have neglected the contributions of Goldstone bosons,
W$^{\pm}$, Z$^{0}$, and $\gamma$ exchange in the quark self-energy.  Of
these, we expect only the Goldstone bosons to have a large effect on the
quark self-energy in the Landau gauge,
and then only in the heavy quark sector where it is
strongly coupled.
Thus the next step in this analysis is to expand our
model to include Goldstone boson exchange.
For heavy quarks, we expect the effects on the quark self-energy from
the electroweak bosons to be significantly smaller than those of the
gluon, Higgs boson, and Goldstone bosons.
\par  We could also expand the flavor sector to include the three-quark-
generation case and possibly the four-quark-generation case.  This would
enable us to calculate the running of the elements of the
three-quark-generation
Cabibbo-Kobayashi-Maskawa matrix and its four-quark-generation analog.
These calculations should be undertaken with due consideration however,
since introducing
additional quark generations significantly increases the length of the
calculations.  Moreover, our two-quark-generation case seemed to  indicate
that the effects on running quark mass functions
due to mixing between generations are negligible.
\par
This method is well-suited to study a variety of
nonperturbative non-Standard-Model
effects.  For example, we could investigate the
case of the third generation
top quark experiencing a new gauge interaction instead of, or
in addition to, the Higgs mechanism
by replacing the Higgs boson exchange with a new gauge boson exchange in the
top quark self-energy.  This would enable us to study models of dynamical
breaking of electroweak symmetry where the top quark plays a special role
\cite{Bardeen,Miransky}.
Alternately, we could study the case of the fourth generation quarks
experiencing
a new gauge interaction, in addition to the Higgs mechanism, which the SM
quarks do not experience, to study new, dynamical effects.
We address these and other, related, points in a future publication.

\subsection*{Acknowledgements}
\par L.L.S. thanks Craig Roberts for discussions and for sponsorship under the
``Thesis Parts" program at Argonne National Laboratory.  D.W.M. thanks the
Department of Physics at U.C. Davis, and especially Ling-Li Chau, Jack
Gunion and Barry Klein, for hospitality during his sabbatical leave while this
work was in progress.  L.L.S. was supported in part by a DOE Traineeship
and by the University of Kansas General Research Fund.   This work was
supported in part by the DOE Grants No.s DE-FG02-85ER-40214 and
DE-FG05-91ER-40636.

\section*{Appendix}
\appendix

Using $g_{\pm}$ we can specify $g_{ij}$:
\begin{equation}
g^{2}_{11} =  g_{+}sin^{2}\theta+g_{-}cos^{2}\theta, \label{g11}
\end{equation}
\begin{equation}
g^{2}_{12}  =  (g_{+}-g_{-})cos\theta \, sin\theta, \label{g12}
\end{equation}
\begin{equation}
g^{2}_{22}  =  g_{+}cos^{2}\theta+g_{-}sin^{2}\theta. \label{g22}
\end{equation}
Then in terms of
\begin{eqnarray}
{\cal A} & = & \frac{d}{y^{2}ln(x_{0}+y)}
-\frac{g^{2}_{11}+g^{2}_{12}}{32\pi^{2}y^{2}}, \\
{\cal B} & = & \frac{g_{12}(g_{11}+g_{22})}{32\pi^{2}y^{2}}, \\
{\cal C} & = & \frac{d}{y^{2}ln(x_{0}+y)}
-\frac{g^{2}_{12}+g^{2}_{22}}{32\pi^{2}y^{2}},
\end{eqnarray}
and
\begin{eqnarray}
{\cal D} & = & (y\Lambda^{2}_{QCD}+M^{2}_{11}+M^{2}_{12})
(y\Lambda^{2}_{QCD}+M^{2}_{12}+M^{2}_{22}) \nonumber \\
 & - & M^{2}_{12}(M_{11}+M_{22})^{2}, \\
{\cal M}_{11} & = & M_{11}(-y \Lambda^{2}_{QCD}-M^{2}_{22})
-M^{2}_{12}M_{22}, \\
{\cal M}_{12} & = & M_{11}M_{12}M_{22}+M_{12}(-y\Lambda^{2}_{QCD}
-M^{2}_{12}), \\
{\cal M}_{22} & = & M^{2}_{12}M_{11}
+M_{22}(-y\Lambda^{2}_{QCD}-M^{2}_{11}),
\end{eqnarray}
we can write the components of the second order differential equation:
\begin{displaymath}
\frac{d^{2}M_{11}}{dy^{2}}=\frac{\Lambda_{QCD}y}{{\cal D}}
({\cal M}_{11}{\cal A}-{\cal M}_{12}{\cal B})
\end{displaymath}
\begin{equation}
+\frac{1}{{\cal AC}-{\cal B}^{2}}
\left( \frac{dM_{11}}{dy}({\cal EC}+{\cal FB})
+\frac{dM_{12}}{dy}({\cal FC}+{\cal GB}) \right), \label{ddm11}
\end{equation}
\begin{displaymath}
\frac{d^{2}M_{12}}{dy^{2}}=\frac{\Lambda_{QCD}y}{{\cal D}}
(-{\cal M}_{11}{\cal B}+{\cal M}_{12}{\cal C})
\end{displaymath}
\begin{equation}
+\frac{1}{{\cal AC}-{\cal B}^{2}}
\left( \frac{dM_{11}}{dy}({\cal EB}+{\cal FA})+\frac{dm_{12}}{dy}
({\cal FB}+{\cal GA}) \right), \label{ddm12}
\end{equation}
\begin{displaymath}
\frac{d^{2}M_{22}}{dy^{2}}=\frac{\Lambda_{QCD}y}{{\cal D}}
(-{\cal M}_{12}{\cal B}+{\cal M}_{22}{\cal C})
\end{displaymath}
\begin{equation}
+\frac{1}{{\cal AC}-{\cal B}^{2}}
\left( \frac{dM_{12}}{dy} ({\cal EB}+{\cal FA})+\frac{dm_{22}}{dy}
({\cal FB}+{\cal GA}) \right), \label{ddm22}
\end{equation}
where
\begin{eqnarray}
{\cal E} & = & -\frac{2{\cal A}}{y}
-\frac{d}{y^{2}ln^{2}(x_{0}+y)\,(x_{0}+y)} \\
{\cal F} & = & \frac{2{\cal B}}{y} \\
{\cal G} & = & -\frac{2{\cal C}}{y}
-\frac{d}{y^{2}ln^{2}(x_{0}+y) \, (x_{0}+y)}.
\end{eqnarray}
The initial condition $M_{1\,ij}$ are given by:
\begin{eqnarray}
\frac{dM_{11}(y)}{dy} |_{y=0}& = & -
\frac{\Lambda^{2}_{QCD}y^{2}}{2 \cdot {\cal D}}
( {\cal M}_{11} {\cal A}-{\cal M}_{12} {\cal B}) = M_{1\,11} \\
\frac{dM_{12}(y)}{dy}|_{y=0} & = & -
\frac{\Lambda^{2}_{QCD}y^{2}}{2 \cdot {\cal D}}
(-{\cal M}_{11} {\cal B}+{\cal M}_{12} {\cal C})=M_{1\,12} \\
\frac{dM_{22}(y)}{dy}|_{y=0} & = & -
\frac{\Lambda^{2}_{QCD}y^{2}}{2\cdot {\cal D}}
(-{\cal M}_{12}{\cal B}+{\cal M}_{22}{\cal C} )=M_{1\,22}.
\end{eqnarray}
\vspace*{0.5cm}
\par
In \cite{Smith} we use $S^{-1}=-i[{\not}{q}A-B]$ and solve the
integro-differential system
\begin{displaymath}
A^{-1}(q^{2})=1+\frac{i}{q^{2}}\int \frac{d^{4}k}{(2\pi)^{4}}
k \cdot q \, g_{Y} [k^{2}-M^{2}(k^{2})]^{-1}\,g_{Y}
\frac{1}{(k-q)^{2}-M^{2}_{H}}
\end{displaymath}
\begin{displaymath}
-\frac{i}{q^{2}} \int \frac{d^{4}k}{(2\pi)^{4}} \left[-(1+\xi)k \cdot q
+2(\xi-1)\frac{q \cdot (k-q)k\cdot (k-q)}{(k-q)^{2}} \right]
\end{displaymath}
\begin{equation}
\times [k^{2}-M^{2}(k^{2})]^{-1} \frac{4\pi \alpha_{s}}{(k-q)^{2}} \label{A}
\end{equation}
\begin{displaymath}
M(q^{2})=M_{0}A^{-1}+i\int \frac{d^{4}k}{(2\pi)^{4}} \, g_{Y}\,
M[k^{2}-M^{2}(k^{2})]^{-1} \, g_{Y} \frac{1}{(k-q)^{2}-M^{2}_{H}}
\end{displaymath}
\begin{equation}
-i(3+\xi)\int \frac{d^{4}k}{(2\pi)^{4}} M[k^{2}-M^{2}(k^{2})]^{-1}
\frac{4\pi \alpha_{s}}{(k-q)^{2}} \ , \label{M}
\end{equation}
where $\xi$ is the gauge parameter. Our gauge choice here is $\xi=0$.
Equations (\ref{A}) and (\ref{M}) are gauge independent at large momenta
where the terms depending on $\xi$ are subleading.
We have solved Eqs. (\ref{A}) and (\ref{M})
by using the angle approximation for the QCD terms, taking $g_{Y}$ as a
constant, and performing the angular integrals exactly for the Higgs
term with a massive Higgs boson.
The result is converted into integro-differential
equations and solved numerically.
The solutions for the mass functions agree very
well with solutions to Eqs. (\ref{ddm11})-(\ref{ddm22}) for a variety of
boundary conditions.  Further applications of the
 $A\neq 1$ solutions
are presented in \cite{Smith}.


\pagebreak
\subsection*{Figure Captions}

{\bf Figure 1}:  $m_{bottom}(y)/\Lambda_{QCD}$ vs. $y$ for $A=5\times 10^{-5}$,
$B=9.77\times 10^{8}$, and inputs $x_{0}=5$, $m_{0}=$4.89, 506 GeV.
The solid (dashed) line is the gluon plus Higgs boson (QCD-only)
interaction result.
\vspace*{.75cm}
\newline {\bf Figure 2}:
$m_{top}(y)/\Lambda_{QCD}$ vs. $y$ for $A=5\times 10^{-5}$,
$B=9.77\times 10^{8}$, and inputs $x_{0}=5$, $m_{0}=$179, 506 GeV.
The solid (dashed) line is the gluon plus Higgs boson (QCD-only) interaction
result.
\vspace*{.75cm}
\newline  {\bf Figure 3}:
$m_{top'}(y)/\Lambda_{QCD}$ vs. $y$ for $A=5\times 10^{-5}$,
$B=9.77\times 10^{8}$, and inputs $x_{0}=5$, $m_{0}=$179, 506 GeV.
The solid (dashed) line is the gluon plus Higgs boson (QCD-only)
interaction result.
\vspace*{.75cm}
\newline {\bf Figure 4}:
$\theta_{cabibbo}(y)$ vs. $y$ for $A=5\times 10^{-5}$,
$B=9.77\times 10^{8}$, and inputs $x_{0}=5$, $m_{0}=$4.89, 506, 179, 506 GeV.
The solid (dashed) line is the gluon plus Higgs boson (QCD-only) interaction
result.

\begin{thebibliography}{Richardson}

\bibitem{CLEO94}  CLEO Collaboration (Jesse Ernst {\em et. al.}), UR-1381A.
Talk given at a 1994 Meeting of the American Physical Society,
Division of Particles
and Fields (DPF94), Albuquerque, NM, 2-6 Aug. 1994.

\bibitem{NJL}  Y. Nambu, G. Jona-Lasinio, {\em Phys. Rev.} {\bf 122}
(1961) 345.

\bibitem{Peskin}  M.E. Peskin, in {\em Recent Advances in Field Theory and
Statistical Mechanics}, eds. J.B. Zuber, R. Stora (Elsevier, 1984).

\bibitem{Fomin} P.I. Fomin, V.P. Gusynin, V.A. Miransky, Yu. A. Sitenko,
{\em Rivista del Nuovo Cimento} {\bf 6}, Series 3, No. 5 (1983).

\bibitem{AJ}  D. Atkinson, P.W. Johnson, {\em Phys. Rev.} {\bf D35}
(1987) 1943; {\em ibid.} {\bf D37} (1988) 2290; {\em ibid.} (1988) 2296.

\bibitem{chiSB}  J.M. Cornwall, R. Jackiw, E. Tomboulis, {\em Phys. Rev.}
{\bf D10} (1974) 2428; \\
V.A. Miransky, P.I. Fomin, {\em Phys. Lett.} {\bf 105B} (1981) 387; \\
D. McKay, H. Munczek, {\em Phys. Rev.} {\bf D42} (1990) 3548.

\bibitem{Pagels76}  H. Pagels, {\em Phys. Rev.} {\bf D14} (1976) 2747; \\
{\em ibid.} {\bf D15} (1977) 2991; \\
J.M. Cornwall, {\em ibid.} {\bf D22} (1980) 1452; \\
J.M. Ball, F. Zachariasen, {\em Phys. Lett.} {\bf 106B} (1981) 133; \\
H.J. Munczek, {\em Phys. Lett.} {\bf B175} (1986) 215.

\bibitem{Pagels79}  H. Pagels, S. Stokar, {\em Phys. Rev.} {\bf D20} (1979)
2947.

\bibitem{Jackiw}  R. Jackiw, K. Johnson, {\em Phys. Rev.} {\bf 8} (1973)
2386.

\bibitem{MM89}  H.J. Munczek, D.W. McKay, {\em Phys. Rev.} {\bf D39} (1989)
888.

\bibitem{Barducci}  A. Barducci, R. Casalbuoni, S. DeCurtis, D. Dominici,
R. Gatto, {\em Phys. Rev.} {\bf D38} (1988) 238.

\bibitem{RW94}  C.D. Roberts, A.G. Williams, {\em Dyson-Schwinger Equations and
their Application to Hadronic Physics}, in {\em Progress in Particle
and Nuclear Physics}, ed. A. Fabler (Pergamon, 1994).

\bibitem{Hadicke}  A. Hadicke, {\em Intern. J. Mod. Phys.} {\bf A6} (1991)
3321.

\bibitem{Roberts93}  C.D. Roberts, {\em Schwinger-Dyson Equations: Dynamical
Chiral Symmetry Breaking and Confinement, in QCD Vacuum Structure}, eds.
H.M. Fried, B. Muller (World Scientific, 1993).

\bibitem{Higash}  K. Higashijima, {\em Phys. Lett.} {\bf B124} (1983) 257;
\\ {\em Phys. Rev.} {\bf D29} (1984) 1228.

\bibitem{RM90}  C.D. Roberts, B.H.J. McKellar, {\em Phys. Rev.} {\bf D41}
(1990) 672.

\bibitem{Kondo}  For a discussion see K. Kondo, H. Nakatani, {\em Mod. Phys.
Lett.} {\bf A4} (1989) 2155.
(1990) 672.

\bibitem{note2}  T. Appelquist {\em et. al.} have determined that next-higher
-order terms beyond the ladder approximation amount to a 1\%-20\% correction,
depending on the fermion representation, for theories in which the coupling is
a slowly running function of momentum and is large enough to trigger
spontaneous
chiral symmetry breaking. \\
T. Appelquist, K. Lane, U. Mahanta, {\em Phys. Rev. Lett.} {\bf 61} (1988)
1553.

\bibitem{Richardson} J.L. Richardson, {\em Phys. Lett.} {\bf B82} (1979)
272.

\bibitem{twoloop}  F.J. Yndurain, {\em Phys. Lett.} {\bf 63B} (1976) 211; \\
E. Reya, {\em Phys. Rep.} {\bf 69} (1981) 195; \\
G. Altarelli, N. Cabibbo, G. Corb\'{o}, L. Maiani, G. Martinelli,
{\em Nucl. Phys.} {\bf B208} (1982) 365.

\bibitem{note3}  We choose a median value, $x_{0}=5$, in comparison with
Munczek and Jain's $x_{0}=10$ \cite{Munczek}, and D. Atkinson {\em et. al.}'s
values $1 \ge x_{0} \ge 7.3$ \cite{AJ}, for example.  We calculate, moreover,
that the running quark mass function does not depend very strongly on $x_{0}$.
The $x_{0}=5$ and $x_{0}=10$ cases vary by 1.6\% and 2.6\%, respectively,
from the $x_{0}=2$ value.

\bibitem{GL}  J. Gasser, H. Leutwyler, {\em Phys. Rep.} {\bf 87} (1982) 77.

\bibitem{Jain94}  P. Jain, A.J. Sommerer, D. W. McKay, J.R. Spence, J.-P.
 Vary, B.-L. Young, {\em Phys. Rev.} {\bf D49} (1994) 2514.

\bibitem{MM90}  D. McKay, H.J. Munczek, {\em Phys. Rev.} {\bf D42} (1990)
3548.

\bibitem{Politzer}  H.D. Politzer, {\em Nucl. Phys.} {\bf B117} (1976) 397.

\bibitem{Munczek}  H.J. Munczek, P. Jain, {\em Phys. Rev.} {\bf D46} (1992)
438.

\bibitem{Koide94} Y. Koide, University of Shizuoko Report No.
US-94-05, October 1994.

\bibitem{PDB94}  Particle Data Group, L. Montanet {\em et. al.}, {\em Phys.
Rev.} {\bf D50} (1994) 1173.

\bibitem{Smith}  L.L. Smith, P. Jain, D.W. McKay, manuscript in preparation.

\bibitem{note1}  We thank Dave Besson for emphasizing this point to us.

\bibitem{Bardeen}  W.A. Bardeen, C.T. Hill, M. Lindner, {\em Phys. Rev.}
{\bf D41} (1990) 1647; \\
Y. Nambu, in {\em New Theories in Physics, Proceedings of the XI International
Symposium on Elementary Particle Physics}, eds. Z. Ajduk, S. Pokorski,
A. Trautman (World Scientific, 1989).

\bibitem{Miransky}  V.A. Miransky, M. Tanabashi, K. Yamawaki, {\em Mod. Phys.
Lett.} {\bf A4} (1989) 1043.

\end{thebibliography}
\end{document}